\documentclass{PoS}

\title{Precise dispersive determination of the f0(600) and f0(980) resonances}

\ShortTitle{Precise dispersive determination of the f0(600) and f0(980) resonances}

\author{\speaker{Jose R. Pel\'aez}%
       \\
       Departamento de F\'{\i}sica Te\'orica II, Universidad Complutense, Spain\\
       E-mail: \email{jrpelaez@fis.ucm.es}}

\author{R. Garc\'{\i}a Mart\'{\i}n\\
        Departamento de F\'{\i}sica Te\'orica II, Universidad Complutense, Spain\\
}

\author{R. Kaminski\\
        Department of Theoretical Physics, H. Niewodnicza\'{n}ski Insitute of
Nuclear Physics, Polish Academy of Sciences, 31-342 Krak\'{o}w, Poland\\
        E-mail: \email{Robert.Kaminski@ifj.edu.pl}
}

\author{J. Ruiz de Elvira\\
        Departamento de F\'{\i}sica Te\'orica II, Universidad Complutense, Spain\\
        E-mail: \email{jacoboruiz@gmail.com}
}

\abstract{We review our recent dispersive and model independent determination of the lightest two scalars, 
in terms of poles and residues -- or mass, width and coupling--by means of once and twice subtracted Roy equations, using as input constrained fits to data, including the most recent ones from Kl4 decays. We find the $f_0(600)$ pole at  $(457^{+14}_{-13})-i(279^{+11}_{-7})$~MeV
and that of the $f_0(980)$ at $(996\pm7)-i(25^{+10}_{-6})$ MeV,
whereas their respective couplings to two pions are $3.59^{+0.11}_{-0.13}$ GeV and 
$2.3\pm0.2$ GeV.}

\FullConference{Sixth International Conference on Quarks and Nuclear Physics,\\
		April 16-20, 2012\\
		Ecole Polytechnique, Palaiseau, Paris}

\begin{document}
\section{Introduction}

The existing experimental
information from $\pi\pi$ scattering has many conflicting data sets at intermediate
energies and no data at all close to the interesting threshold region. For many years this
fact has made it very hard to obtain conclusive results on $\pi\pi$ scattering at low energies
or in the $\sigma$ (or $f_0(600)$) and $f_0(980)$ region. However, recent \cite{:2009pv} and precise experiments on kaon decays,
related to $\pi\pi$ scattering at very low energies, have renewed the interest on
this process.

The dispersive integral formalism is model independent, just based on analyticity
and crossing, and relates the $\pi\pi$ amplitude at a given energy with an integral over the
whole energy range, increasing the precision and providing information on the amplitude
at energies where data are poor, or where there is no data, like the complex plane. In addition, it makes the
parametrization of the data irrelevant once it is included in the integral and relates
different scattering channels among themselves.

Roy equations (RE), based on twice subtracted
dispersion relations and crossing symmetry
conditions for $\pi\pi\rightarrow \pi\pi$ amplitudes were
obtained in 1971 \cite{Roy:1971tc}. In recent years, these equations
have been used either to obtain predictions
for low energy $\pi\pi$ scattering, either using
Chiral Perturbation Theory (ChPT)\cite{Ananthanarayan:2000ht, Colangelo:2001df}, or to
test ChPT \cite{DescotesGenon:2001tn,Kaminski:2006qe,Kaminski:2006yv}, as well as to solve old data ambiguities
\cite{Kaminski:2002pe}. The RE are relevant for the sigma
pole, whose position has also been predicted very
precisely with the help of ChPT \cite{Caprini:2005zr}.
Our group \cite{Kaminski:2006qe,Kaminski:2006yv} has also used RE with
Forward Dispersion Relations (FDR) to obtain
a precise determination of $\pi\pi$ scattering amplitudes
from data consistent with analyticity, unitarity
and crossing. On purpose, we have not
included ChPT constraints, so that we can use
our results as tests of the ChPT predictions. Unfortunately,
the large experimental error of the
scattering length $a_0^2$ of the isospin 2 scalar partial
wave, becomes a very large error for the sigma
pole determination using RE. For this reason,
a new set of once-subtracted RE, called GKPY
eqs. for brevity, have been derived \cite{GarciaMartin:2011cn}. Both the RE
and GKPY equations provide analytic extensions
for the calculation of poles in the complex plane.
Actually, we review here our recent results \cite{GarciaMartin:2011cn} on simple data fits constrained to satisfy
these dispersive representations as well as our results \cite{GarciaMartin:2011jx}
for the $\sigma$ and $f_0(980)$ poles
in the S0 wave obtained from GKPY eqs.

\section{Overview of the analysis}

The approach we have followed throughout a series of
works~\cite{Kaminski:2006qe,Kaminski:2006yv,GarciaMartin:2011cn,Pelaez:2004vs} can be summarized as follows:

 First, we obtain simple fits to data for each $\pi\pi$ scattering
partial wave (the so called {\em Unconstrained Fits to Data}, or {\em UFD}
for short). These fits are uncorrelated, therefore they can be very easily
changed when new, more precise data become available. Let us remark
that for our latest results we have used previous fits for all
waves except the S0 wave, that we improve in \cite{GarciaMartin:2011cn}.
For this wave, below 850 MeV, we have included the very precise
$K_{l4}$ data~\cite{:2009pv}, we got rid of the controversial $K\rightarrow \pi\pi$ point,
and we have included the isospin correction to $K_{l4}$ data from 
\cite{Colangelo:2008sm}.
Above 850 MeV we have updated the S0 wave using a polynomial fit to improve
the intermediate matching between parametrizations (with a continuous derivative)
and the flexibility of the $f_0(980)$ region, which is of particular importance for the discussion of the ``dip'' and ``no-dip'' scenarios that we will comment below..
At different stages of our approach we have also fitted
Regge theory \cite{Pelaez:2003ky} to $\pi\pi$ high energy data, and as our precision
was improving, we have improved some 
of the UFD fits with more flexible parametrizations.

Then, these UFD are checked against FDR, several sum rules, RE and GKPY eqs.
The UFD fit does not satisfy very well these dispersion relations. Particularly the GKPY eqs. for the S0 wave in the $f_0(980)$ region are
satisfied very poorly \cite{GarciaMartin:2011cn}. 

Finally, we impose these dispersive relations in the previous
fits as additional constraints. These new {\em Constrained Fits
to Data} ({\em CFD} for short) are much more precise and reliable
than the UFD set, being consistent with analyticity, unitarity, crossing,
etc. The price to pay is that now all the waves are correlated.

In order to quantify how well the dispersion relations are satisfied,
we define six quantities $\Delta_i$ as the difference between the left and right
sides of each dispersion relation
whose uncertainties we call $\delta\Delta_i$.
Next, we define the average discrepancies
\begin{equation}
\bar{d}_i^2\equiv\frac{1}{\hbox{number of points}}
\sum_n\left(\frac{\Delta_i(s_n)}{\delta\Delta_i(s_n)}\right)^2,
\end{equation}
where the values of $s_n$ are taken at intervals of 25 MeV. 
Note the similarity with an averaged $\chi^2/(d.o.f)$
and thus $\bar{d}_i^2\leq1$
implies fulfillment of the corresponding dispersion relation within errors.
In Table 1 we show the average discrepancies of the UFD 
for each FDR eq. up to 1420 MeV, and for each RE and GKPY eq. up to 1100 MeV. 
Since the total average discrepancies lie between 1 and 1.6 standard deviations, they can be
clearly improved by imposing simultaneous fulfillment of dispersion
relations. This is actually done 
in the CFD set, which is obtained by minimizing:
\begin{eqnarray}
\chi^2&\equiv
\left\{\bar{d}^2_{00}+\bar{d}^2_{0+}+\bar{d}^2_{I_t=1}+\bar{d}^2_{S0}+\bar{d}^2_{S2}+\bar{d}^2_P\right\}W\\
&+\bar{d}^2_I+\bar{d}^2_J+\sum_i\left(\frac{p_i-p_i^{\rm exp}}{\delta p_i}\right)^2,
\end{eqnarray}
where $p_i^{exp}$ are all the parameters of the different UFD
parametrizations for each wave or Regge trajectory, 
thus ensuring the data description, and
$d_I$ and $d_J$ are the discrepancies for a couple of crossing
sum rules, see reference~\cite{GarciaMartin:2011cn} for details. 
Note that we choose $W\simeq 9-12$ for the effective number of degrees of freedom needed to parametrize curves like those appearing in the S0, P and S2 waves.

From Table~\ref{data} it is clear that the CFD set satisfies remarkably 
well all dispersion relations within uncertainties, and 
hence can be used directly if one needs a consistent parametrization.
But, in addition, it can be used inside certain sum rules to obtain precise
predictions for the threshold parameters of the effective range expansion for $\pi\pi$ scattering
\cite{GarciaMartin:2011cn,Nebreda:2012ve}. At least
for the scattering lengths and slopes, they are remarkably compatible with the prediction of
\cite{Bern}, and seem to be easily accommodated within two-loop ChPT \cite{Nebreda:2012ve}. However, the description of the shape parameters (third order in the effective range expansion) seem to call for even higher orders of ChPT \cite{Nebreda:2012ve}.

A relevant remark about the use of GKPY Eqs. is that it has allowed us \cite{GarciaMartin:2011cn} to settle a longstanding controversy between a ``dip'' and ``non-dip'' solution for the inelasticity of the S0 wave right above the $K\bar K$ threshold. The ``dip'' solution is clearly favored by the GKPY Eqs., and this is of relevance for the precise determination of the $f_0(980)$ resonance properties from scattering data.

In summary the CFD set provides a model independent and very precise
description
of the $\pi\pi$ scattering data consistent with analyticity and crossing.

\begin{table}[h]
\begin{tabular}{l|c||c}
&(UFD)&(CFD)\\\hline
&$s^{1/2}\leq 1420\,$MeV&$s^{1/2}\leq 1420\,$MeV\\
$\pi^0\pi^0$ FDR& 2.13 & 0.51\\
$\pi^+\pi^0$ FDR& 1.11 & 0.43\\
$I_{t=1}$ FDR& 2.69 & 0.25\\\hline
&$s^{1/2}\leq 1100\,$MeV&$s^{1/2}\leq 1100\,$MeV\\
Roy eq. S0 & 0.56&0.04\\
Roy eq. S2 & 1.37&0.26\\
Roy eq. P  & 0.69&0.12\\\hline 
GKPY eq. S0 & 2.42&0.24\\
GKPY eq. S2 & 1.14&0.11\\
GKPY eq. P  & 2.13&0.60\\
\end{tabular}
\caption{Average discrepancies $\bar d^2$ of the UFD and CFD for each 
FDR and RE.  Note the remarkable CFD consistency.}\label{data}
\end{table}

\section{Position of the $\sigma$ and $f_0(980)$ poles}

The mass and width of the $\sigma$ meson quoted in the
Particle Data Table are very widely spread~\cite{PDG06}
\begin{equation}
M_{\sigma} - i\frac{\Gamma_{\sigma}}{2} \approx (400 - 1200) - i(250 -
500)
\mbox{\boldmath{MeV}}.
\end{equation}
The main reason of these uncertainties is that $\pi\pi$ scattering data
 are few and sometimes contradictory.
Moreover, all quoted theoretical models are 
not equally reliable, and less so when extending the amplitude
to the complex plane.
Thus the position of the sigma 
pole in various models differ significantly~\cite{PDG06}, although,
with a couple of exceptions, they tend to agree roughly around $\sim (450\pm 50) -i\, (250\pm50)$, particularly those results based on dispersion theory.

The mass and width of the $f_0(980)$
meson quoted in the
Particle Data Table are~\cite{PDG06}

\begin{equation}
M_{f_0(980)} - i\frac{\Gamma_{f_0(980)}}{2} \approx (970 - 990) - i(20 -
50)
\mbox{\boldmath{MeV}}.
\end{equation}

The recent data from E865 collaboration at Brookhaven~\cite{Pislak2001} and from NA48/2~\cite{:2009pv} 
provide us with new and very precise information on the $\pi\pi$ scattering at low energies.
Thanks to these new data we are 
able to construct, with our Constrained Fits to Data,
a very reliable description
for the $S0$ wave especially near the $\pi\pi$ threshold.

With those precise data parametrizations,
we can now use either RE or GKPY eqs. to
extend the partial waves analytically to the complex plane 
and look for poles in the second sheet of the S-matrix. 
As it is well known, a pole on the second 
Riemann sheet (unphysical sheet) is associated with a
zero on the first---the physical one.


Depending on whether we use Roy or GKPY Eqs. we find a different accuracy in
our results, namely:
\begin{eqnarray}
\hspace*{-1cm}&\sqrt{s_{\sigma}} =  (445 \pm 25) - i\,(278^{+22}_{-18}) \mbox{ \boldmath{MeV}} & {\rm (RE)} 
\label{sigmaRE} \\
\hspace*{-1cm}&\sqrt{s_{\sigma}} =  (457^{+14}_{-13}) - i\,(279^{+11}_{-7}) \mbox{ \boldmath{MeV}} 
& {\rm (GKPY)}
\label{sigmaGKPY}  
\end{eqnarray}
and for the $f_0(980)$ pole:
\begin{eqnarray}
\hspace*{-1cm}&\sqrt{s_{f_0(980)}} =  (1003^{+5}_{-27}) - i\,(21^{+10}_{-8}) \mbox{ \boldmath{MeV}}  & {\rm (RE)} 
\label{sigmaRE} \\
\hspace*{-1cm}&\sqrt{s_{f_0(980)}} =  (996\pm 7) - i\,(25^{+10}_{-6}) \mbox{ \boldmath{MeV}} 
& {\rm (GKPY)}.
\label{sigmaGKPY}  
\end{eqnarray}
These values are in good agreement with each other. Note that for the $f_0(980)$ we have had to add a 4 MeV systematic uncertainty on the imaginary part of the pole position, which comes as the difference between using our isospin symmetric formalism with the charged or the neutral kaon mass.

In the case $\sigma$, on the one hand, both the mass and width
lie less than 1 standard deviation away 
from the prediction of 
twice-subtracted RE combined with ChPT results
for the scattering lengths \cite{Caprini:2005zr}: 
$\sqrt{s_{\sigma}} = 441^{+16}_{-8} -i\,272^{+9}_{-14.5} 
\mbox{\boldmath{MeV}}$.
On the other hand our pole determination above
is roughly two standard deviations 
from the mass and width in our simple fit of a conformal expansion to low energy data 
\cite{GarciaMartin:2008cq} $\sqrt{s_{\sigma}} = (484\pm17) -i\,(255\pm10) 
\,\mbox{\boldmath{MeV}}$.

In the case of the $f_0(980)$, the mass is somewhat higher than that quoted
in the PDG $980\pm10\,$ MeV, although note that ours is the pole position and is model independent. Concerning the width, which once again we obtain from 
the pole position as $\Gamma=-2 Im \sqrt{s_{f_0(980)}}=50^{+20}_{-12}\,$MeV it lies within the range given in the PDG, namely, $40-100\,$ MeV.

\section{Conclusions}

The GKPY equations \cite{GarciaMartin:2008cq,Kaminski:2008rh}---Roy-like dispersion relations with one subtraction 
for the $\pi\pi$ amplitudes---provide stringent
constraints for dispersive analysis of experimental data.
We have provided simple and ready to use parametrizations, constrained to satisfy these equations as well as Roy equations and froward dispersion relations, that simultaneously describe the existing data.

The main advantage of GKPY eqs. is that, for the same input,
 in the $0.45 \mbox{ GeV}\,\leq\sqrt{s}\leq 1.1\mbox{ GeV}$
region they have  significantly smaller errors than standard Roy. eqs.
Hence, they provide better accuracy tests 
and analytic extensions of the amplitudes in that region.
In particular, using just a data analysis consistent within errors 
with Forward Dispersion Relations, Roy eqs. and GKPY eqs. (and no ChPT input),
we have presented here our recent very precise determination
of the $\sigma$ pole position: 
\begin{equation}
\sqrt{s_{\sigma}} = (457^{+14}_{-13}) - i\,(279^{+11}_{-7}) \mbox{ \boldmath{MeV}},
\end{equation}
and of the $f_0(980)$:
\begin{equation}
\sqrt{s_{f_0(980)}} = (996\pm 7) - i\,(25^{+10}_{-6})  \mbox{ \boldmath{MeV}}.
\end{equation}


\section{Acknowledgements}
Work supported by Spanish contract 
FPA2011-27853-C02-02 and Polish Ministry of Science and
Higher Education (grant No N N202 101 368). We
 acknowledge support from the European Community-Research Infrastructure Integrating Activity Study of Strongly
InteractingMatter (HadronPhysics2, Grant n. 227431) under the EU 7th Framework Programme. 



\bibliographystyle{aipproc}   

\begin{thebibliography}{99}
\vspace*{-0.25cm}


\bibitem{:2009pv}
Precise tests of low energy QCD from Ke4 decay properties.
NA48 collaboration. CERN-PH-EP-2010-036.
   J.~R.~Batley {\it et al.}  [NA48/2 Collaboration],
   Phys.\ Lett.\  B {\bf 677}, 246 (2009).

\bibitem{Roy:1971tc}
  S.~M.~Roy,
  Phys.\ Lett.\  B {\bf 36}, 353 (1971).

\bibitem{Ananthanarayan:2000ht}
  B.~Ananthanarayan, G.~Colangelo, J.~Gasser and H.~Leutwyler,
  Phys.\ Rept.\  {\bf 353}, 207 (2001).

\bibitem{Colangelo:2001df}
  G.~Colangelo, J.~Gasser and H.~Leutwyler,
  Nucl.\ Phys.\  B {\bf 603}, 125 (2001).

\bibitem{DescotesGenon:2001tn}
  S.~Descotes-Genon, N.~H.~Fuchs, L.~Girlanda and J.~Stern,
  Eur.\ Phys.\ J.\  C {\bf 24}, 469 (2002).

\bibitem{Kaminski:2006qe}
  R.~Kaminski, J.~R.~Pelaez and F.~J.~Yndurain,
  Phys.\ Rev.\  D {\bf 77}, 054015 (2008).

\bibitem{Kaminski:2006yv}
  R.~Kaminski, J.~R.~Pelaez and F.~J.~Yndurain,
  Phys.\ Rev.\  D {\bf 74}, 014001 (2006)
  [Erratum-ibid.\  D {\bf 74}, 079903 (2006)].

\bibitem{Kaminski:2002pe}
  R.~Kaminski, L.~Lesniak and B.~Loiseau,
  Phys.\ Lett.\  B {\bf 551}, 241 (2003).

\bibitem{Caprini:2005zr}
  I.~Caprini, G.~Colangelo and H.~Leutwyler,
  Phys.\ Rev.\ Lett.\  {\bf 96}, 132001 (2006).

\bibitem{GarciaMartin:2011cn} 
  R.~Garcia-Martin, R.~Kaminski, J.~R.~Pelaez, J.~Ruiz de Elvira and F.~J.~Yndurain,
  Phys.\ Rev.\ D {\bf 83}, 074004 (2011).

\bibitem{GarciaMartin:2011jx} 
  R.~Garcia-Martin, R.~Kaminski, J.~R.~Pelaez and J.~Ruiz de Elvira,
  Phys.\ Rev.\ Lett.\  {\bf 107}, 072001 (2011).

\bibitem{Pelaez:2004vs}
  J.~R.~Pelaez and F.~J.~Yndurain,
  Phys.\ Rev.\  D {\bf 71}, 074016 (2005).

\bibitem{Colangelo:2008sm}
  G.~Colangelo, J.~Gasser and A.~Rusetsky,
  Eur.\ Phys.\ J.\  C {\bf 59}, 777 (2009).

\bibitem{Pelaez:2003ky}
  J.~R.~Pelaez and F.~J.~Yndurain,
  Phys.\ Rev.\  D {\bf 69}, 114001 (2004).


\bibitem{Nebreda:2012ve} 
  J.~Nebreda, J.~R.~Pelaez and G.~Rios,
  arXiv:1205.4129 [hep-ph].

\bibitem{Bern}
  G.~Colangelo, J.~Gasser and H.~Leutwyler,
  Nucl.\ Phys.\  B {\bf 603}, 125 (2001).
  B.~Ananthanarayan, G.~Colangelo, J.~Gasser and H.~Leutwyler,
  Phys.\ Rept.\  {\bf 353}, 207 (2001).



\bibitem{PDG06} W. M. Yao {\it et al.}, J. Phys. G33, 1-1232 (2006).

\bibitem{Pislak2001} S. Pislak {\it et al.}, Phys. Rev. Lett. 87, 221801 (2001).

\bibitem{Yndurain:2007qm}
  F.~J.~Yndurain, R.~Garcia-Martin and J.~R.~Pelaez,
  Phys.\ Rev.\  D {\bf 76}, 074034 (2007).

  
\bibitem{GarciaMartin:2008cq}
  R.~Garcia-Martin, R.~Kaminski and J.~R.~Pelaez,
  Int.\ J.\ Mod.\ Phys.\  A {\bf 24}, 590 (2009).

\bibitem{Kaminski:2008rh}
  R.~Kaminski, R.~Garcia-Martin, P.~Grynkiewicz and J.~R.~Pelaez,
  Nucl.\ Phys.\ Proc.\ Suppl.\  {\bf 186}, 318 (2009).
  R.~Garcia-Martin, R.~Kaminski and J.~R.~Pelaez,
  Int.\ J.\ Mod.\ Phys.\  A {\bf 24}, 590 (2009).

\end{thebibliography}


\end{document}